\documentclass[twocolumn,letterpaper]{IEEEAerospaceCLS}  

\usepackage[]{graphicx}    
\newcommand{\ignore}[1]{}  
\usepackage[hidelinks]{hyperref}
\usepackage{astjnlabbrev}
\usepackage{multicol}

\begin{document}


\newcommand{\vismin}{0.38}
\newcommand{\vismax}{0.75}
\newcommand{\nirmin}{0.87}
\newcommand{\nirmax}{1.63}

\title{The Pandora SmallSat: A Low-Cost, High Impact Mission to Study Exoplanets and Their Host Stars}

\author{%
Thomas Barclay, Elisa V. Quintana, Knicole Colón, Benjamin J. Hord,\\Gregory Mosby, Joshua E. Schlieder, Robert T. Zellem\\ 
NASA Goddard Space Flight Center\\
8800 Greenbelt RD.\\
Greenbelt, MD 20771\\
thomas.barclay@nasa.gov
\and
Jordan Karburn, Lance M. Simms, Peter F. Heatwole\\ 
Lawrence Livermore National Laboratory\\
7000 East Ave.\\
Livermore, CA 94550\\
karburn1@llnl.gov
\and
Christina L. Hedges\\
University of Maryland, Baltimore County\\
1000 Hilltop Circle\\
Baltimore, MD 21250\\
christina.l.hedges@nasa.gov
\and
Jessie L. Dotson, Thomas P. Greene\\
NASA Ames Research Center\\
Moffett Field, CA 94035\\
jessie.dotson@nasa.gov
\and
Trevor O. Foote, Nikole K. Lewis\\
Cornell University\\
Ithaca, NY 14853\\
tof2@cornell.edu
\and
Benjamin V. Rackham\\
Massachusetts Institute of Technology\\
Cambridge, MA 02139\\
brackham@mit.edu
\and
Brett M. Morris\\
Space Telescope Science Institute\\
Baltimore, MD\\
bmmorris@stsci.edu
\and
Emily A. Gilbert\\
Jet Propulsion Laboratory,\\
California Institute of Technology\\
Pasadena, CA\\
emily.a.gilbert@jpl.nasa.gov
\and
Veselin B. Kostov\\
SETI Institute\\
339 Bernardo Ave. Suite 200\\
Mountain View, CA\\
veselin.b.kostov@nasa.gov
\and
Jason F. Rowe\\
Bishop's University\\
2600 College\\
Sherbrooke, QC\\
jason.rowe@ubishops.ca
\and
Lindsay S. Wiser\\
Arizona State University\\
781 Terrace Mall\\
Tempe, AZ 85287\\
lindsey.wiser@asu.edu
\and
Dániel Apai\\
University of Arizona\\
933 North Cherry Avenue\\
Tucson, AZ 85721
\thanks{\footnotesize 979-8-3503-5597-0/25/$\$31.00$ \copyright2025 IEEE}              
}

\maketitle

\thispagestyle{plain}
\pagestyle{plain}

\maketitle

\thispagestyle{plain}
\pagestyle{plain}

\begin{abstract}

The Pandora SmallSat is a NASA flight project aimed at studying the atmospheres of exoplanets—planets orbiting stars outside our Solar System. Pandora will provide the first dataset of simultaneous, multiband (visible and NIR), long-baseline observations of exoplanets and their host stars. Pandora is an ambitious project that will fly a 0.44~m telescope in a small form factor. To achieve the scientific goals, the mission requires a departure from the traditional cost-schedule paradigm of half-meter-class observatories. Pandora achieves this by leveraging existing capabilities that necessitate minimal engineering development, disruptive and agile management, trusted partnerships with vendors, and strong support from the lead institutions. The Pandora team has developed a suite of high-fidelity parameterized simulation and modeling tools to estimate the performance of both imaging channels. This has enabled a unique bottom-up approach to deriving trades and system requirements. Pandora is a partnership between NASA and Lawrence Livermore National Laboratory. The project completed its Critical Design Review in October 2023 and is slated for launch into Sun-synchronous, low-Earth orbit in Fall 2025.

\end{abstract}

\tableofcontents

\section{Introduction}
The Pandora SmallSat mission --- selected in 2021 in NASA's inaugural Astrophysics Pioneers program --- is designed to robustly characterize the atmospheres of exoplanets---planets orbiting distant stars---using the transmission spectroscopy technique. This technique works by observing planets as they pass in front of their host stars and obtaining high-precision spectra of the starlight that filters through a planet's atmosphere. Each planet produces a unique spectral fingerprint that can be used to infer the makeup and composition of its atmosphere. NASA's Hubble Space Telescope (HST) and James Webb Space Telescope (JWST) have successfully used this technique to characterize the atmospheres of many dozens of planets, however, the use of this technique to confidently detect and characterize the atmospheres of planets as small as Earth has remained elusive due to challenges imposed by the host stars. A major limiting factor in obtaining these high-precision measurements is that many stars are magnetically active, with bright and dark star spots that induce spectral contamination. Pandora is designed to address stellar contamination by collecting long-duration observations, with simultaneous visible and short-wave infrared wavelengths, of exoplanets and their host stars. These data will help us understand how contaminating stellar spectral signals affect our interpretation of the chemical compositions of planetary atmospheres. Over its one-year prime mission, Pandora will observe more than 200 transits from at least 20 exoplanets that range in size from Earth-size to Jupiter-size, and provide a legacy dataset of the first long-baseline visible photometry and near-infrared (NIR) spectroscopy catalog of exoplanets and their host stars \cite{Quintana2024}. 

NASA developed the Astrophysics Pioneers program to enable compelling science in a new regime of cost (\$20M cost cap) and scope that is higher than CubeSats but lower than Explorers missions, with a mission lifecycle of 5 years. The Pioneers program is categorized under NASA's Research and Technology Projects (which adhere to rules set in NASA's Procedural Requirements (NPR) 7120.8, albeit with additional requirements and reviews) rather than standard NASA spaceflight programs (which adhere to NPR 7120.5). This translates to lighter-touch management and oversight from NASA, and less rigid requirements, than what is expected for standard spaceflight programs --- all of which are beneficial for adhering to the five-year complete mission duration. There are also expectations for what must be accomplished within the mission, and they include building and testing the observatory, launch, operations, data processing, data archiving, and publishing results to demonstrate the science objectives have been accomplished. 

Pioneers missions are by nature high-risk programs. Unique and novel planning and solutions --- from both the programmatic side and the technical side --- throughout the project's lifecycle are required to accomplish mission objectives. Agile management and strong partnerships with frequent communication between scientists, engineers, managers, vendors, partners, and stakeholders has been critical to ensure milestones are met within schedule and cost restraints. The communication loop between scientists and engineers, including detector and hardware testing inputs and outputs and data simulations, has proven valuable to drive design trades and system requirements in a nimble and low cost manner. 

The Pioneers program is designed to provide leadership experience to early career scientists and engineers with a goal of broadening and diversifying the next generation of space mission leaders. To meet these goals, Pandora has developed a mentoring team structure \cite{Dotson2024} where early career scientists and engineers hold many of the key leadership roles, and more senior team members provide support in Deputy positions (this also helps with minimizing costs).

The factors that make the Pandora mission unique include: flying a half-meter-class telescope in a SmallSat; a payload with two detector channels that enable simultaneous visible photometry and NIR spectroscopy time series; and the long-baseline observations that can be obtained for a given target (which is difficult for larger missions due to the high-cost and highly oversubscribed nature of these facilities).

Pandora is a partnership between NASA, Lawrence Livermore National Laboratory (LLNL), the University of Arizona (for mission operations), and over a dozen institutions. LLNL holds a key role in managing the mission, including handling all major procurements and leading the integration and testing. Pandora recently passed its Critical Design Review in October 2023, and has a scheduled launch readiness in Fall 2025.






\section{Science Objectives}
One of NASA's priority science goals for the coming decades is to identify habitable Earth-like worlds in other planetary systems \cite{Astro2020}. To this end, NASA has a number of current and upcoming facilities designed to detect and characterize the atmospheres of distant exoplanets with the goal of searching for life beyond our solar system. NASA's most recently launched Astrophysics Flagship mission, JWST, has capabilities to probe the atmospheres of transiting exoplanets as small as Earth to high precision using the transmission spectroscopy technique. This technique works by obtaining spectra of a star while a planet transits (as viewed from the observer) and comparing it to spectra when the planet is out of transit, the difference yielding the spectra due to the starlight that filters through a planet's atmosphere. This starlight can be absorbed by some molecular species in the planetary atmosphere and transmitted by others, and the amount of absorption/transmission for a given planet is wavelength dependent. For a given wavelength, greater absorption yields a more opaque atmosphere, and thus a deeper measured planetary transit (inferring a greater planet size). The absorption spectrum of our own planet is primarily influenced by water, carbon dioxide, and molecular oxygen absorption, whereas Venus's atmosphere is primarily composed of carbon dioxide, which creates a thicker and denser atmosphere compared to Earth. Every planet provides a unique spectral fingerprint, and by modeling this transmission spectrum with atmospheric composition models, we can determine the chemical makeup and structure of a distant planet's atmosphere.

A challenge that arises is that brightness variations due to spots on a star's surface can induce spectral contamination if not accurately accounted for. This can lead to signatures in the data that can mimic or mask features in the planetary atmosphere \cite{Rackham2018,Rackham2023}. The challenge of identifying and mitigating stellar contamination in transmission spectroscopy measurements has been recognized for over a decade \cite{McCullough2014,Rackham2017,Zellem2017,Rackham2018,Apai2018,Espinoza2019,Rackham2019b,Rackham2019a,Wakeford2019,Barclay2021}. Early results from JWST have shown stellar contamination to be a limiting factor in the definitive identification of atmospheres on Earth-size planets \cite{Lim2023,Moran2023}. Extrapolation of both planetary atmosphere models and stellar contamination models illustrate that shorter wavelength data (blueward of JWST's shortest wavelength of 0.6 micron) can help break the degeneracies (see also e.g., \cite{Rackham2023}).

Pandora's primary goal is to address stellar contamination by utilizing long-baseline visible photometry ($\vismin$--$\vismax$ $\mu$m) and simultaneous NIR spectroscopy ($\nirmin$--$\nirmax$ $\mu$m) to disentangle star and planet signals in transmission spectra to reliably determine exoplanet atmosphere compositions. Pandora has two primary objectives. The first is to determine the spot covering fractions of low-mass exoplanet host stars and the impact of these active regions on exoplanetary transmission spectra. Pandora's visible channel will measure brightness over time that can vary as star spots rotate into and out of view. Pandora's NIR channel, which will operate simultaneously with the visible channel, will measure how the spectral features vary with stellar rotation. This multiwavelength data will enable a thorough analysis of the stellar spectrum such that the spot, faculae, and quiescent photosphere coverage can be decomposed \cite{Zhang2018,Wakeford2019,Narrett2024} and a number of various modeling and retrieval methods have been developed to use such data to help quantify and correct for stellar contamination \cite{Rackham2018,Wakeford2019,Iyer2020,Rackham2023}. The transmission spectra from Pandora is obtained in a wavelength region where water is a known strong molecular absorber. Once the data is corrected for stellar contamination, Pandora's second objective is to robustly identify exoplanets with hydrogen- or water-dominated atmospheres, and determine which planets are likely covered by clouds and hazes. 

Pandora will operate concurrently with JWST, complementing JWST's deep-dive, high-precision spectroscopy measurements with broad wavelength, long-baseline observations. Pandora's science objectives are well-suited for a SmallSat platform and illustrate how small missions can be used to truly maximize the science from larger flagship missions. 

To meet the mission science objectives, Pandora will select 20 primary science exoplanet host stars and collect data of a minimum of 10 transits per target, with each observation lasting about 24 hours. This results in 200 days of science observations required to meet mission requirements. Pandora has a primary mission lifetime of one year, and to optimize the observations the Pandora team has developed a scheduler tool \cite{Foote2023}. Pandora will have capabilities (and observing time beyond that needed for the primary science goals) to support additional science cases during the primary mission. The Pandora Science Team has established an Auxiliary Science Working Group to identify feasible science cases, and plans to engage the external community are under development.



Pandora's science return will also be enhanced with ground-based multi-band photometry. These observations started $\sim$1~year prior to Pandora's launch and will be used to provide ephemeris refinement and to further constrain the long-term variability of each Pandora target's host star. These data will be combined with data collected from the Pandora spacecraft to model impacts of stellar variability on transit observations. Ground-based data is being obtained both by professional observatories (including the Las Cumbres Observatory \cite{Brown2013}) and amateur astronomers and citizen scientists. Pandora is collaborating with Exoplanet Watch a citizen science initiative to observe transiting exoplanets with small ground-based telescopes \cite{Zellem2020}.

\section{Technical Design Overview}

\subsection{The Pandora Observatory}
The Pandora observatory includes a payload developed by LLNL and a spacecraft bus delivered by Blue Canyon Technologies (BCT). The payload consist of an all-aluminum 45~cm Cassegrain telescope with a large outer baffle, providing off-axis light suppression to maximize the field of regard. An optical relay, cooling system, and two detector assemblies (for the visible and NIR channels) reside behind the telescope. The optical design splits the light via a dichroic, and includes a relayed visible path and an infrared path that is spectrally dispersed by a prism. Visible photometry ($\vismin$--$\vismax$ $\mu$m) and NIR slitless-spectroscopy ($\nirmin$--$\nirmax$ $\mu$m) will be obtained simultaneously for each Pandora target. The system is designed to maximize throughput and stability.

\begin{figure}
\centering
\includegraphics[width=0.45\textwidth]{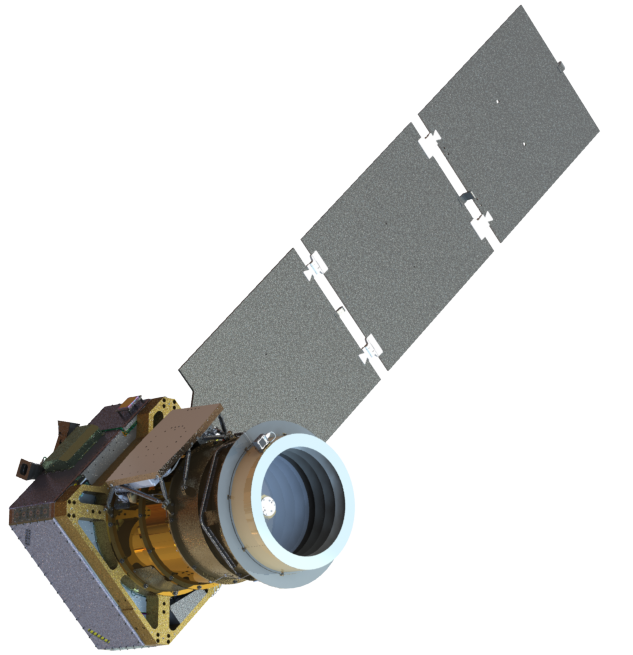}
\caption{The Pandora observatory shown with the solar array deployed. Pandora is designed to be launched as a ride-share attached to an ESPA Grande ring. Very little customization was carried out on the major hardware components of the mission such as the telescope and spacecraft bus. This enabled the mission to minimize non-recurring engineering costs.}
\label{fig:pandora}
\end{figure}

Power is provided by a single wing solar array that can be driven in one axis. Pandora is designed to be flown as a ride-share attached to an ESPA Grande ring and the solar array is deployed after separation from the faring. 

\subsection{Pandora Telescope}
\begin{figure}
\centering
\includegraphics[width=0.35\textwidth]{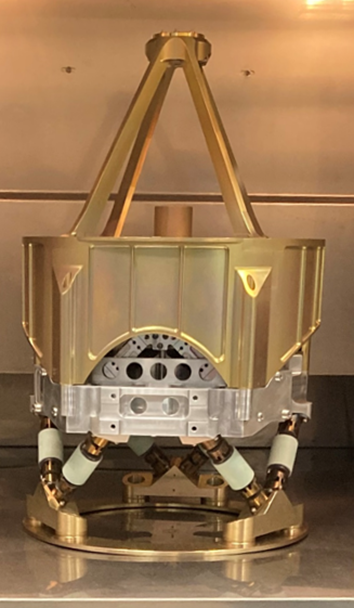}
\caption{An engineering development unit of the Pandora telescope. This half-meter class telescope enables Pandora to collect the exquisite data needed to measure the atmospheres of distant exoplanets.}
\label{fig:coda}
\end{figure}

\begin{figure}
\centering
\includegraphics[width=0.35\textwidth]{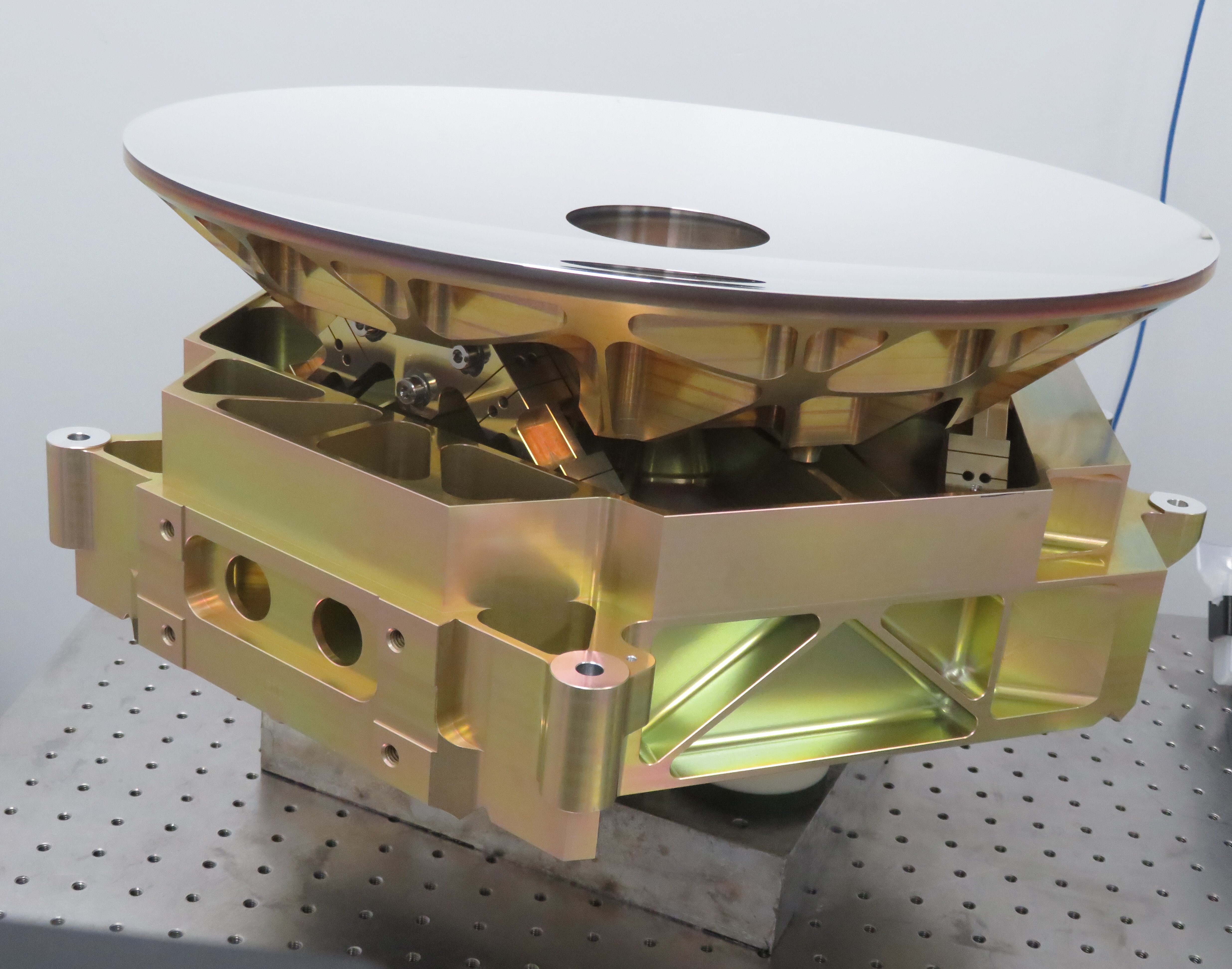}
\caption{The polished and coated flight unit primary mirror of the Pandora telescope. The mirror is mounted on bipods, secured to the optical bench.}
\label{fig:primary}
\end{figure}

At the heart of Pandora is the 44~cm Cassegrain telescope (an engineering design unit is shown in Figure~\ref{fig:coda}) that will provide the large number of photons necessary to meet the tight precision requirements on both the visible and infrared channels and is key to enabling Pandora's exoplanet science. The primary mirror is a paraboloid (shown in Figure~\ref{fig:primary}) and the secondary mirror is a 9.2~cm diameter hyperboloid supported by a four-bar spider. The all-aluminum telescope was designed by LLNL and manufactured by Corning Specialty Materials. The optical surface is coated in silver providing high reflectivity through visible and infrared wavelengths.

\subsection{Visible Detector Assembly (VDA)}
The Visible Detector Assembly (VDA) includes an Excelitas pco.panda 4.2 back illuminated scientific CMOS camera and a thermoelectric cooler. The visible channel serves two purposes --- it provides short-wavelength data required to meet the exoplanet science objectives, and it enables the payload to compute attitude errors for inclusion in the attitude control loop, thus serving a role as a fine guidance sensor.

Visible light is directed through the optical relay onto the sensor, providing an effective field of view of a circle with 0.3 degrees diameter. The field of view was selected to enable viewing of the target star and sufficient stars to calculate an astrometric solution. 

The sensor has low read noise of less than 2 electrons per read, which is important as we will run the sensor at 5 Hz. The relatively rapid readout speed is required by the attitude control system. The pixel size of 6.5 $\mu$m results in an angular size on the sky of 0.78 arcsec/pix. 

The data volume from storing full frame images at 5 Hz time resolution would rapidly exceed our downlink capacity. To reduce data volume: (1) we select 5--9 sources, the science target plus a number of nearby stars, from which to store the data during each science observation, and save a 50$\times$50 pixel region centered on the targets; and (2) during science observations we plan to sum together 50 images, storing data at 10 s cadence. This time resolution is short enough that we do not expect any astrophysical changes in the sources over this time, and spacecraft pointing drift should also be minimal. 

\begin{figure*}
\centering
\includegraphics[width=0.95\textwidth]{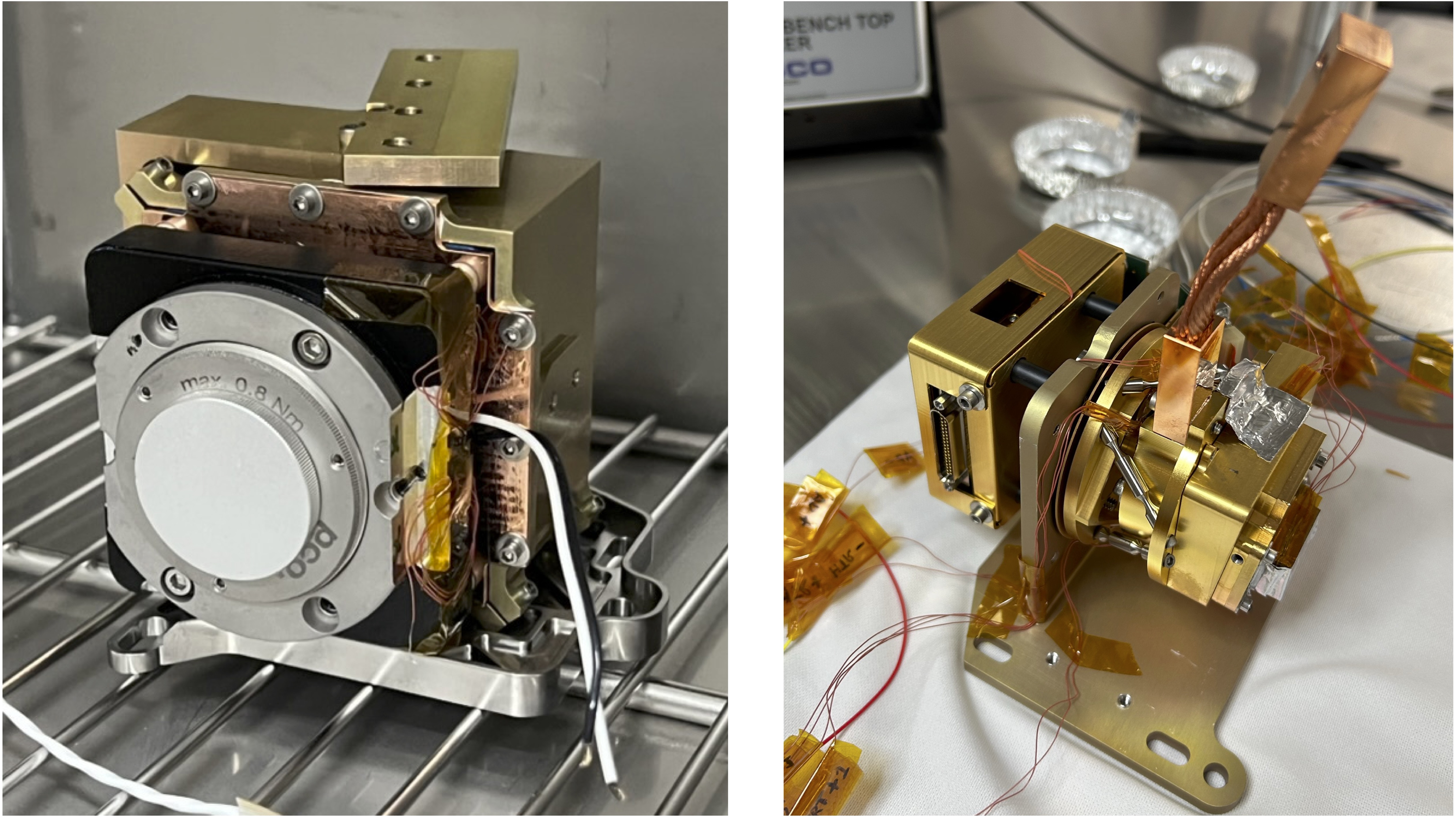}
\caption{Assembled Pandora detector assemblies. These are the flight assemblies. The left image is the VDA and the right is the NIRDA. These two detectors enable Pandora to collect contemporaneous visible and infrared light observations of exoplanet host stars.}
\label{fig:detectors}
\end{figure*}

The VDA completed environmental tests that raise its technical readiness level to 6.

\subsection{Near Infra-Red Detector Assembly (NIRDA)}
The Near Infra-Red Detector Assembly (NIRDA) includes a Teledyne HAWAII-2RG (H2RG) sensor and SIDECAR ASIC, both flight spares from the JWST Near Infrared Camera (NIRCam) instrument \cite{Rieke2005}. The NIRDA is controlled via a Markury Scientific MACIE controller card. A region of approximately 1300$\times$600 pixels is exposed to light from the sky with the remaining covered, providing baffling from stray light and thermal background. There is a snout that extends a few centimeters from the detector to further reduce the level of thermal background reaching the sensor. Below the snout, but just above the sensor is a thermal cutoff filter that provides out of band blocking at an optical density of 5. This is necessary because the sensor is sensitive to light out to 2.5 $\mu$m but it would be blinded by thermal emission from the optics without a cutoff filter. The blue edge of the filter was set at 0.87 $\mu$m to limit the length of the infrared spectrum to help reduce the potential for overlap of the science target spectrum with spectra of closeby astrophysical sources.

Temperature stability is key to meeting the exoplanet science requirements. A temperature below 110 K is maintained to a stability of better than 10 mK on the flight sensor via a thermal control system that uses a cryocooler. 

The sensor is read non-destructively, and in science operations will operate in subarray mode. Even though the sensor has a size of 2048$\times$2048 pixels we plan to typically only read a 400$\times$80 pixel subarray. This will yield a frame time of 0.32 s. We plan to typically collect 24 non-destructive reads before each sensor reset. This provides us with a science spectrum every 8 s. We only store two resultant frames per integration to minimize the data that needs to be downlinked. Each of those resultant frames will be the average of 4 frames; we will store an average of the first four and an average of the last four frames in each integration.

\subsection{Integration and Test of the Observatory}
The observatory bus and payload are being integrated separately. The bus is being integrated by BCT in Colorado, and the camera assemblies are being integrated at LLNL and are going through an extensive testing campaign. After completion of the detector testing, both flight units will be shipped to Corning for integration with the telescope. Payload system assembly, integration, and testing (AI\&T) will be performed at LLNL. When complete, the payload will ship to BCT's facilities for final AI\&T.

\begin{figure*}
\centering
\includegraphics[width=0.95\textwidth]{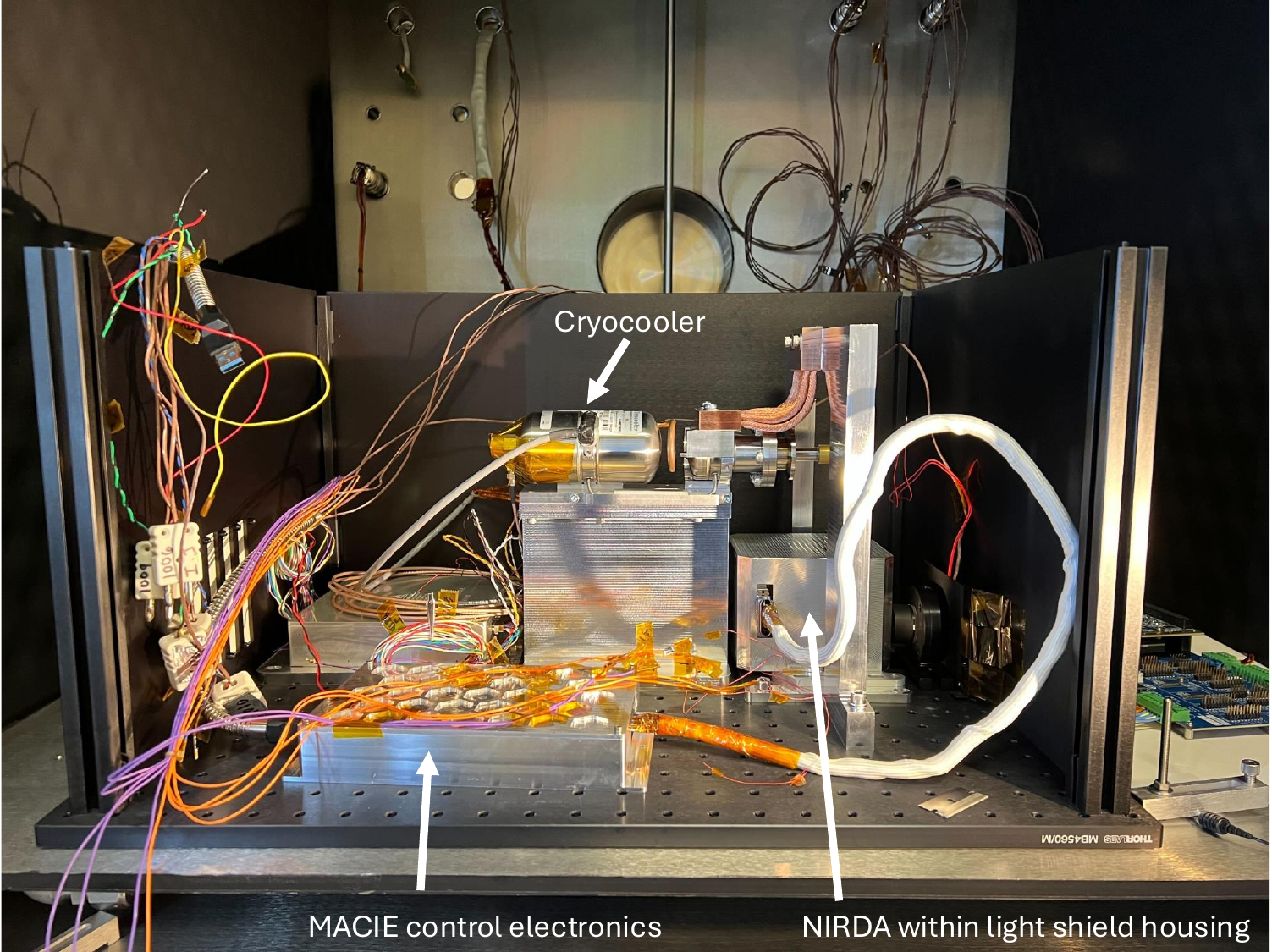}
\caption{A testing setup for NIRDA characterization testing inside the TVAC chamber at LLNL. This testing program was used to both verify the performance requirements, and instrument flight software. The test also contributed important calibration reference data products.}
\label{fig:detectorstesting}
\end{figure*}

As shown in Figure~\ref{fig:detectorstesting}, the two detector assemblies, thermal control system, and all payload electronics were demonstrated end-to-end during several thermal vacuum campaigns \cite{Foote2024}. These data are used to verify that flight software meets requirements, and also to generate calibration products that will be used in the Pandora data analysis pipeline. These products include measurements of dark and bias, read noise, detector gain, saturation, linearity, flat fields, bad pixels, and persistence. 

\subsection{Concept of Operations and Observation Planning}
Pandora will be launched into a Sun-synchronous, low-Earth orbit with an altitude of approximately 600\,km and a longitude of ascending node of approximately 6am or 6pm. This enables access to the whole sky over the year of science operations. Following a one month checkout and commissioning period, there will be a one-year prime mission during which the science objectives need to be met. An extended mission phase is possible following the prime mission.

To meet science objectives, Pandora needs to observe at least 20 exoplanet host star ten times each, for a total of 200 transit observations. Science activities are broken up into a visit, which is defined as the period of time where observations of one science target are prioritized \cite{Foote2023}. Visits typically last for 24 hours, although they can be much shorter where gaps in the schedule needed to be filled. These visits are then further subdivided into observation sequences during which the boresight of the telescope is fixed on the single sky locations. During an observation sequence (which could last more than an hour) the spacecraft boresight will be inertially fixed, with no roll needed. There is a small continuous viewing region during each Pandora visit, but for most science targets the spacecraft will need to slew away when the target is occulted by the Earth. During those times, we point the spacecraft at a secondary target that is selected to enable ancillary science or for payload performance monitoring activities. Most visits consist of 15 orbits around the Earth with two observation sequences per visit.

Pandora will utilize the KSAT-lite network of ground stations. Data are planned to be downlinked to Earth via Pandora's X-band high gain antenna up to four times per day. These downlinks are planned to occur outside times of critical science observations. Spacecraft commands are uplinked weekly via S-band.

\section{Data Simulations}

An essential component of the implementation of the Pandora mission has been the use of high-fidelity simulated science data products. Trade studies needed to be performed during the initial formulation of the mission, and so the science team worked closely with the engineering team to develop simulated science data products to address these trades. The simulation software has continued to be developed \cite{Foote2022,Hedges2024} and is in use daily to refine requirements and to make decisions when hardware issues arise.

\subsection{Simulations for optical performance trade studies}

A key requirement that needed to be set early in mission development was that of image quality due to the large impact this could have on the cost of the optical system. However, tracing requirements on image quality, such as point spread function (PSF) diameter or encircled energy, to telescope parameters was not trivial. To rapidly iterate between potential designs and requirements the optical engineering team developed the framework to generate PSFs based upon optical designs. These models were then provided to the science team who simulated observatory performance and reported back to the engineering team whether these performance models met science requirements. The science team were able to show the impacts of engineering decisions on Pandora's high level science requirements. This ultimately led to confidence that Pandora could be built with minimal modifications to an existing telescope design, which was originally developed by LLNL for a different program.

\begin{figure*}
\centering
\includegraphics[width=0.95\textwidth]{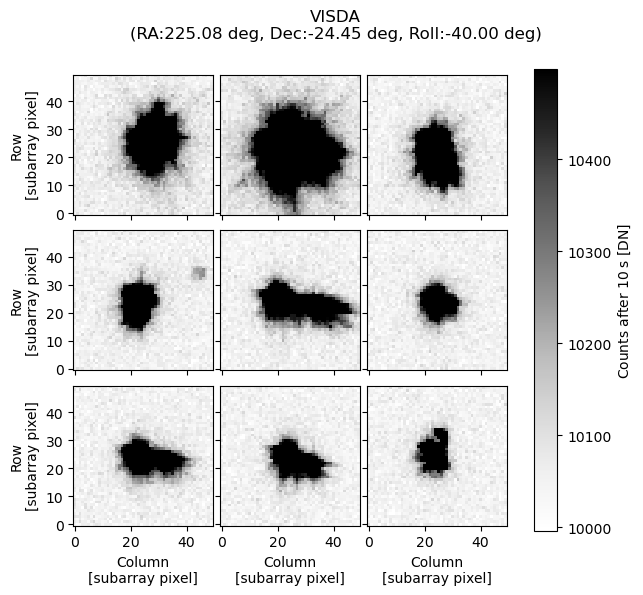}
\caption{Simulated science images were generated using projected observatory performance data to verify that Pandora would meet science requirements. The panels show nine targets centered in 50$\times$50 pixel windows, representative of the expected downlinked data for a single 10 s integration. The target in the top left is the science target, and the remaining 8 sources are used for comparison in the science analysis. This data uses as-built models and measured detector performance.}
\label{fig:visperformance}
\end{figure*}

\begin{figure}
\centering
\includegraphics[width=0.45\textwidth]{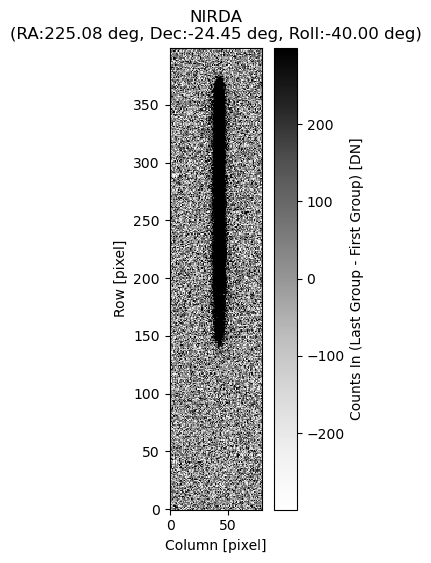}
\caption{Synthetic data from the infrared channel generated by our simulation software performance. These simulations include realistic jitter performance at sub-second high time resolution. This image is from an 8 s integration time. We expect to collect thousands of such images each day during science operations.
}
\label{fig:irperformance}
\end{figure}

As mission formulation advanced, a qualification unit telescope was fabricated for early testing, which included pre-environmental tests. Infrared wavefront errors were estimated from measured visible light data to estimate system performance as a function of wavelength and field angle. These were used to generate PSFs for the visible and infrared channels and incorporated into the Pandora simulation software. Figures~\ref{fig:visperformance} and \ref{fig:irperformance} show example output from our simulation software.

Later in the mission development it became clear that meeting our image quality requirements may be challenging. Fortunately we were able to assess what level of degraded performance we could deal with, and crucially identify whether there were other aspects of the observatory performance where we could enhance performance and trade this at the expense of image quality. We were able to improve our stray light rejection through a change to the baffle design which enabled us to accept a larger PSF knowing that our background light level was lower than designed.

\subsection{Verifying pointing stability}

A key technical performance metric for Pandora is pointing stability. Achieving sub-arcsecond level control is key to increasing the viability of small satellites for numerous astronomical science cases, but doing this in a small payload is often challenging. 

The BCT Saturn bus has the capability to accept pointing error estimations from the Pandora payload to increase pointing stability and the technology to achieve fine pointing control has been demonstrated on previous smallsats, such as Asteria \cite{Knapp2020}, however it was important to the Pandora mission to demonstrate that that the flight software we developed would work on realistic data.

\begin{figure}
\centering
\includegraphics[width=0.45\textwidth]{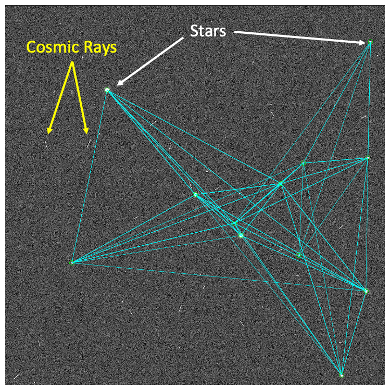}
\caption{High-fidelity simulations have enabled us to demonstrate the performance of our astrometric solution generation software. This image shows a single 0.2 s duration frame from the visible camera. The image contains a realistic star field along with cosmic rays and detector noise. The flight software detects the stars and computes attitude errors which are then used by the spacecraft bus in the attitude control system.}
\label{fig:pointing}
\end{figure}

During each 5\,Hz acquisition of science data on the VDA camera, each image is used to calculate pointing. The science team generated simulated images from the visible camera with a realistic astronomical scene and realistic detector effects. The simulations included spacecraft boresight jitter, image noise, and cosmic rays. A synthetic 200\,ms image is shown in Figure~\ref{fig:pointing}, from which the payload determines a right ascension, declination, and rotation for the sensor. The payload calculates a quaternion that represents the rotation from the detector frame to the Earth-centered inertial (ECI) frame. The payload then computes a quaternion that describes the commanded quaternion with respect to the measured quaternion. This attitude error quaternion can then be converted to three-axis attitude errors which are delivered to the BCT bus at a 5\,Hz rate.

Using these simulated data, we were able to verify the payload could detect the stars, generate an astrometric solution, and compute spacecraft attitude errors using the spacecraft flight electronics, even in the presence of expected jitter. Moreover, we demonstrated that this could be done on a timescale that would allow these attitude errors to be provided to the bus at 5 Hz. This enabled us to retire a significant program risk. Having access to simulations of the science performance early in the mission was very useful. This enabled good communication between the science and engineering teams on what goals we were working towards.

\section{Lessons Learned}
Pandora was selected in the first round of the NASA Astrophysics Pioneers program. Pioneers is responsive to the NASA Small Spacecraft Strategic Plan\footnote{\url{https://www.nasa.gov/wp-content/uploads/2020/03/smallsatstrategicplan.pdf?emrc=d9a523}}, which provides recommendations for NASA to sustain the integration of small missions (particularly ESPA-Class platforms, like Pandora, that can achieve more complex science than CubeSats) into a balanced portfolio of flight systems. Many of the proposed strategies within this Strategic Plan, especially those that support disruptive technology innovation, are strategies that Pandora has adopted that have been critical to the success so far. These include embracing the use of standardized spacecraft and spacecraft parts to reduce risk, and avoiding mission-specific solutions where possible. Pandora procured a number of Commercial Off-the-Shelf (COTS) parts that in some cases included more capabilities than were needed, but overall were a cost savings. Pandora was also able to acquire Government-furnished equipment, including the telescope that leveraged investments made by another government agency and the JWST NIRCam flight spare detector (as well as the assistance from members of the NIRCam team to understand how to test it). Full institutional commitments from the lead institutions for help with both resource identification and problem-solving has been a key factor in Pandora's success so far. 

For Pandora, partnering with another government institution (Department of Energy/LLNL) that has extensive experience with small spacecraft, and having that partner manage the mission, has enabled NASA to leverage expertise, processes, and hardware that would otherwise have been cost and/or schedule prohibitive. Embracing agile practices has also been useful, such as the Project Management's use of an agile approach towards requirements development versus adopting a typical rigid structure that is often used by larger NASA missions. Developing trusted relationships with vendors is important, as this helped with applying the ``80/20 rule'', in which an 80\% solution is targeted for a capability at 20\% of the traditional cost. Large firm fixed price contracts have also been helpful to stay within the mission cost cap.

On the technical side, frequent communication among the scientists and engineers, with a liaison who is familiar with the jargon and culture from each side, proved to be valuable. The communication loop between test engineers and the science team enabled a smooth path towards identifying trade studies and defining requirements. 

Developing simulations of the science performance to check that scientific requirements are being met, and starting this effort early in mission development, was useful. This enabled good communication between science and engineering to identify the goals we were all working towards. 

In addition to providing a platform to enable compelling science at low cost and mission duration, a goal of the Pioneers program is to provide leadership and mission experience for new Principal Investigators and early career scientists and engineers \cite{Dotson2024}. The Pandora mission team is therefore composed of many team members learning for the first time how to navigate the development and implementation of space flight projects, all within a newly developed program with a scope that has not been done before. To make this work, the team also includes members with extensive experience in space mission design, development, and management. We developed a mentoring plan to match up early career members with a more senior team member, and this has been highly successful to both the mission and individual career development. In addition to career levels, the Pandora science team is also diverse on other axes, and the early development of a Code of Conduct has facilitated a healthy environment that leads to a highly-engaged and productive team. 

The Pandora team has also implemented frequent meetings that have evolved according to the needs within each development stage, but a daily meeting with mission leadership (with the PI, Deputy PI, Project Manager (PM), Deputy PM, Project Scientist (PS), and Deputy PS) has ensured that the mission stays on track and nothing falls through the cracks. 

Some challenges remain, and may take time to evolve, such as the procurement mechanisms (some contracts can take over six months to establish, which can lead to a mission risk when trying to complete a mission within five years). Additional lessons learned are sure to arise as the mission heads towards launch, operations, and closeout. Documenting what worked and what did not work is an important part of the Pioneers program, and is necessary to further the development of NASA's small space missions portfolio.

\section{Conclusions}

NASA's Astrophysics Pioneers program offers a great platform for high-impact science at low cost and quick turnaround. Pandora's primary science goals --- to quantify and correct for stellar contamination in space-based exoplanet transmission spectroscopy --- will address a science area that fills a gap in NASA's Astrophysics roadmap and addresses a problem that has recently gained more attention amongst the science community due to recent high-profile results from JWST. Specifically, a number of JWST observing programs aimed at detecting and characterizing atmospheres on Earth-like worlds are finding that stellar spectral contamination is plaguing their results. Typical transmission spectroscopic observations for exoplanets from large missions like JWST focus on collecting data for one or a small number of transits for a given target, with short observing durations before and after the transit event. In contrast to large flagship missions, SmallSat platform enable long-duration measurements for a given target. Pandora can thus collect an abundance of out-of-transit data that will help characterize the host star and directly address the problem of stellar contamination. The Pandora Science Team will select 20 primary science exoplanet host stars that span a range of stellar spectral types and planet sizes, and will collect a minimum of 10 transits per target, with each observation lasting about 24 hours. This results in 200 days of science observations required to meet mission requirements. With a one-year primary mission lifetime, this leaves a significant fraction of the year of science operations that can be used for spacecraft monitoring and additional science. 

The Pandora observatory includes a BCT spacecraft bus and a payload developed by LLNL that includes an all-aluminum 45~cm Cassegrain telescope. Two detector assemblies reside behind the telescope for the visible and NIR channels. The Visible Detector Assembly (VDA) will be used to collect high-precision photometry, and the Near Infra-Red Detector Assembly (NIRDA) will collect simultaneous time-series spectroscopy. The Pandora has been developing high-fidelity simulated science data products that have been valuable tools to converge on science requirements. For example, the engineering team was able to generate PSFs for a given optical design, enabling the science team to simulate the observatory performance to check if requirements were being met. A strong and frequent communication loop between the engineers and scientists proved valuable to many facets of the mission development.

Pandora's success so far can also be attributed to: strong support from all lead institutions; adopting agile practices from the flourishing SmallSat industry; and diverse science and engineering teams with early career participation in key leadership roles. Documenting lessons learned are an important part of sustaining a new program like the Pioneers. 

Pandora has a scheduled launch readiness in the Fall of 2025. The mission will be operational alongside NASA's other flagships, and opportunities for simultaneous observing across facilities will maximize the science even further. Pandora will be capable of an extended mission, should support be available

\acknowledgments
Pandora is supported by NASA’s Astrophysics Pioneers Program. The Pandora team is grateful to the Small Satellite and Special Projects Office at Wallops Flight Facility/Goddard Space Flight Center and the Small Spacecraft Systems Virtual Institute (S3VI) for providing support and resources that have been critical to Pandora's success. The authors also thank Bruce Yost, Roland Vanderspek, Bernard Rauscher, and many others who contributed to Pandora reviews that improved many facets of the mission. 


\bibliographystyle{IEEEtran}

\bibliography{pandora}




\thebiography
\begin{biographywithpic}
{Thomas Barclay}{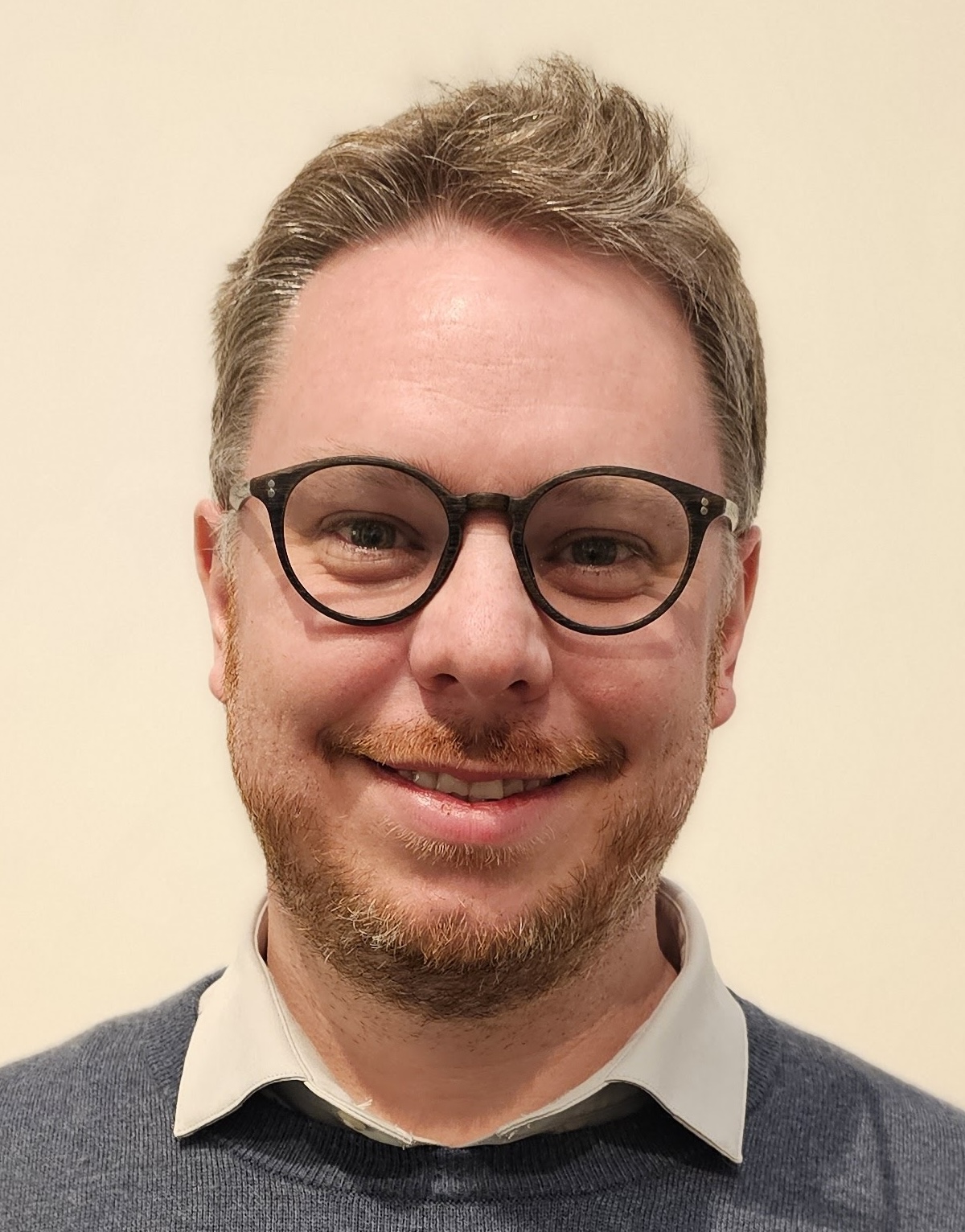}
is an Astrophysicist at NASA Goddard Space Flight Center where he is the Deputy Project Scientist of Pandora and Operations Project Scientist for the Nancy Grace Roman Space Telescope. He previously served in scientific leadership roles for the Kepler and TESS mission. He analyzes data from both ground and space-based exoplanet surveys to search for small, rocky planets and he specializes in devising new methods to confirm their planetary nature. He received his Ph.D. from the University College London in 2011. He has participated in the discovery of over 800 exoplanets and is known for his discovery of Kepler-37b, a planet about the size of the Moon that is the smallest planet known outside of our Solar System. 
\end{biographywithpic} 

\begin{biographywithpic}
{Elisa Quintana}{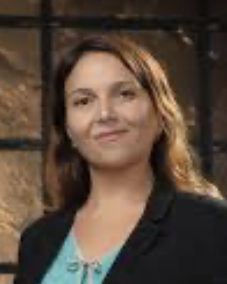}
is an Astrophysicist at NASA’s Goddard Space Flight Center. She is the Principal Investigator of the Pandora SmallSat mission which seeks to characterize the atmospheres of planets orbiting distant stars, with a focus on measuring star spots and their impact on exoplanet atmosphere measurements. She has been working on discovering planets around other stars since she joined the Kepler Mission at NASA Ames Research Center in 2006. She also builds computer models to study how planets form in extreme environments. She is best known for leading a team of astronomers to confirm Kepler-186f, the first Earth-sized planet found to orbit another star in a region that could potentially host extraterrestrial life. She received a Ph.D. in Physics from the University of Michigan, Ann Arbor, in 2004, and joined Goddard in 2017. 
\end{biographywithpic} 

\begin{biographywithpic}
    {Knicole Colón}{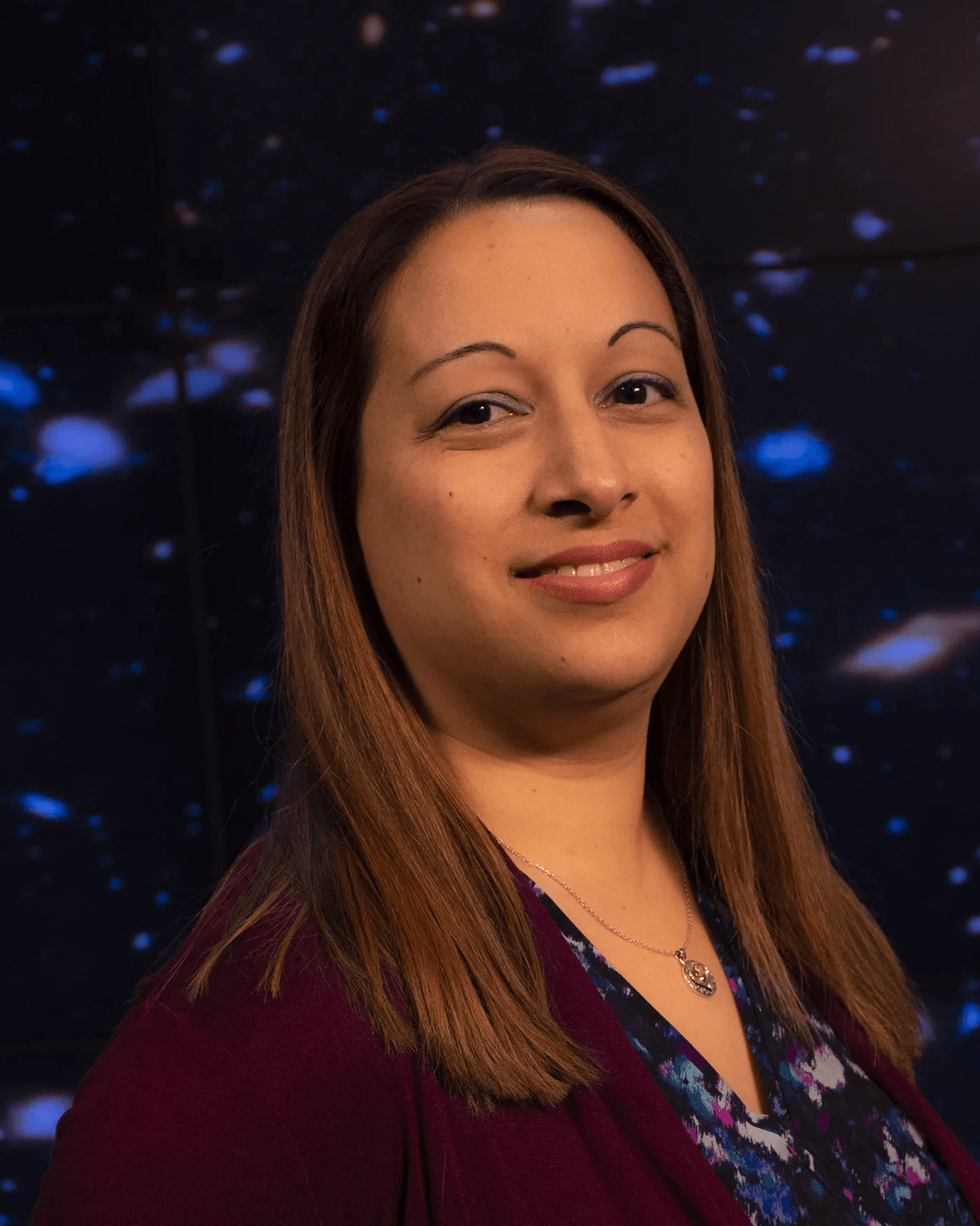}
    is a research astrophysicist based at NASA’s Goddard Space Flight Center. She has held various roles on NASA missions over the years, most recently serving as the Pandora Project Scientist, JWST Operations Project Scientist, and TESS Project Scientist. Her research interests include the study of extreme exoplanets, like the super puffy planet KELT-11b, the disintegrating planet K2-22b, and planets on highly eccentric orbits like HD 80606b. Her experience includes using optical and infrared and ground and space facilities to study such exoplanets.
\end{biographywithpic}

\begin{biographywithpic}
{Benjamin Hord}{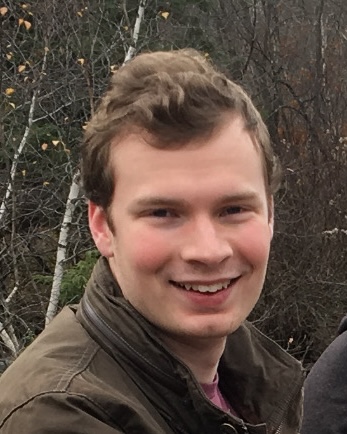} is a NASA Postdoctoral Program fellow at NASA Goddard Space Flight Center. He is a member of Pandora's Science Operation Center, Data Processing Center, and Science Team responsible for the development of software tools, science target selection, simulation of data, and other tasks critical to the functioning of science operations. His current research interests include the formation, evolution, and system architectures of giant planets as well as planet atmosphere observability using a variety of space- and ground-based facilities. Prior to his current position, he earned BAs in Astrophysics and History from Columbia University and a PhD in Astronomy from the University of Maryland.
\end{biographywithpic}

\begin{biographywithpic}
{Gregory Mosby}{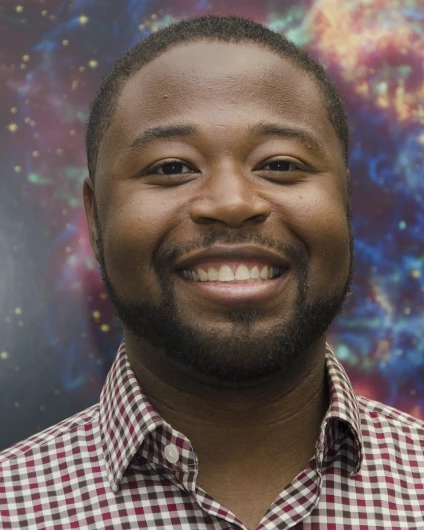} graduated with an Astronomy \& Physics B.S. in 2009 from Yale University. In 2016, he finished his doctoral studies on analyzing low signal-to-noise spectra of quasar host galaxies and near infrared detector optimization at the University of Wisconsin-Madison. His current research interests include near infrared detectors, astronomical instrumentation, and applications of machine learning to observational astronomy. He is a project science team member for the Nancy Grace Roman Space Telescope and is the detector scientist for the Pandora Mission.
\end{biographywithpic}

\begin{biographywithpic}
{Joshua Schlieder}{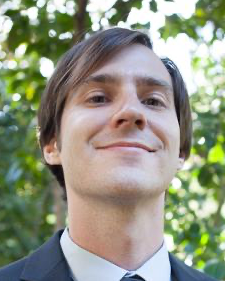} is an Astrophysicist in the Exoplanets and Stellar Astrophysics Lab at NASA's Goddard Space Flight Center in Greenbelt, MD. He is the Wide Field Instrument Scientist for the Nancy Grace Roman Space Telescope, the Operations Project Scientist for the Neil Gehrels Swift Observatory, and a member of the Pandora team. Schlieder’s research focuses on low-mass stars and their exoplanets. This includes discovery and characterization using wide-field, time-domain survey telescopes and narrow-field imaging and spectroscopy with observatories like the JWST. Schlieder received a Ph.D. in Physics from Stony Brook University and performed postdoctoral research at The Max Planck Institute for Astronomy and NASA's Ames Research Center. He also previously held a staff position at the NASA Exoplanet Science Institute at Caltech/IPAC.
\end{biographywithpic}

\begin{biographywithpic}
{Robert Zellem}{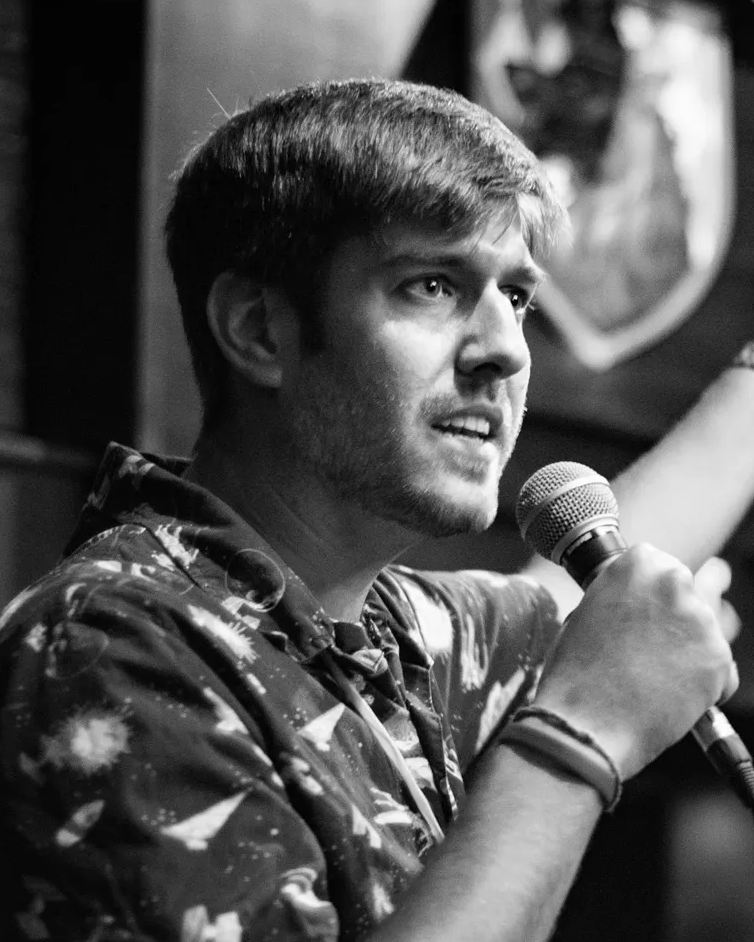} is an astrophysicist at NASA’s Goddard Space Flight Center. His research focuses on the characterization of the atmospheres of exoplanets using both the transit and direct imaging methods. He is the Deputy Project Scientist for Communications for NASA’s Nancy Grace Roman Space Telescope where he is the primary liaison between the Roman Project Science team and Goddard's Office of Communications. He is also a member of the Roman Coronagraph Project Science team where he led the development of the science calibration plan. He is the Project Scientist and Lead for Exoplanet Watch, a citizen science project to observe transiting exoplanets to update their ephemerides to ensure the efficiency use of large telescope time. He is the Ground-Based Sub-working Group co-lead for Pandora, whereby he is coordinating ground-based observations to support both the operations and scientific interpretation of Pandora data. He is also a co-lead for NASA’s Nexus for Exoplanet System Science (NExSS) and a science team member for the NASA’s CASE contribution to ESA’s Ariel mission.
\end{biographywithpic}

\begin{biographywithpic}
{Jordan Karburn}{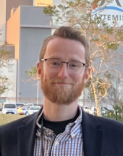} joined LLNL's National Security Engineering Division in 2017 after earning his Bachelor of Science in Electrical Engineering from California Polytechnic State University, San Luis Obispo. He works on flight hardware development projects within the Space Program. He is the Deputy Project Manager for the NASA Pandora SmallSat. In addition to his role on Pandora, he is the Principal Investigator for the CODA telescope program and serves as the Group Leader for Space Electronic Systems within LLNL Engineering Directorate.
\end{biographywithpic} 

\begin{biographywithpic}
    {Lance Simms}{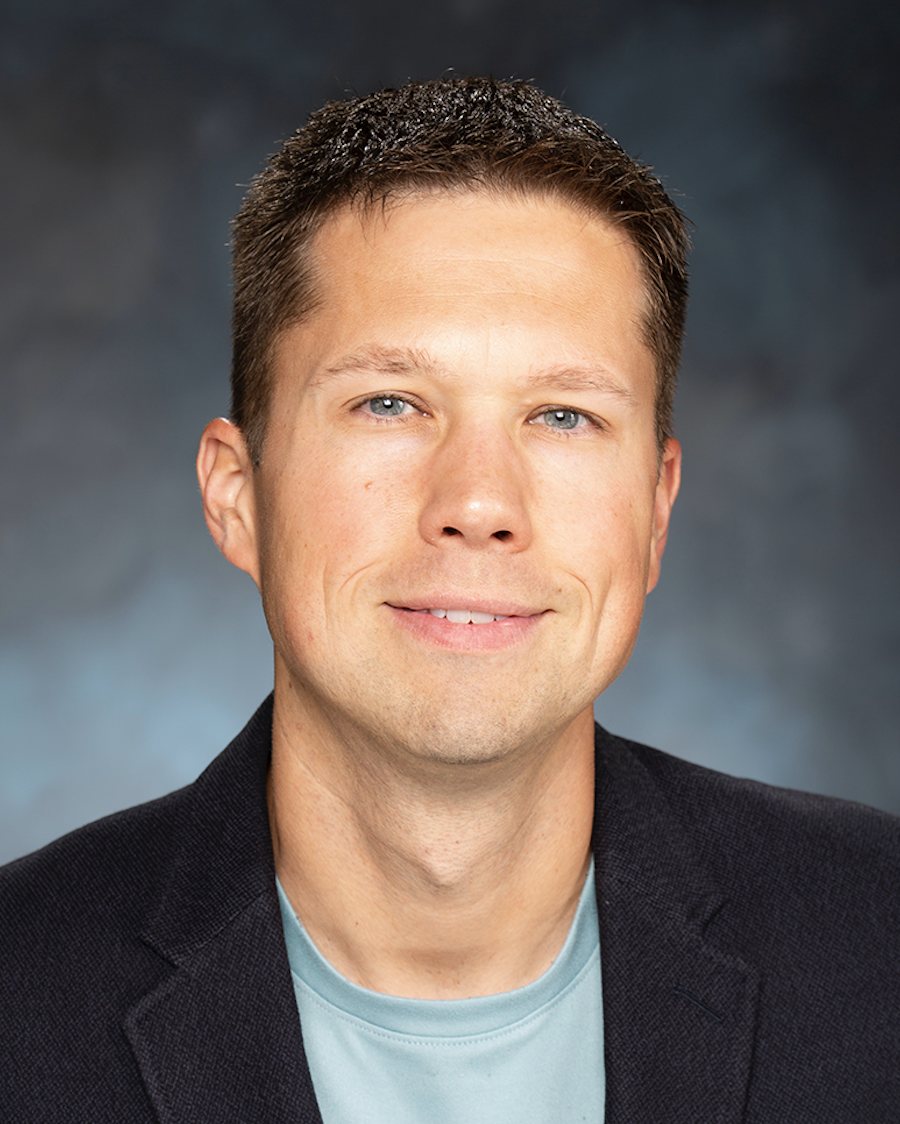} grew up in the suburbs of Chicago and developed an early passion for physics and astronomy. After earning a scholarship from the Adler Planetarium, he pursued his BS in Physics at the University of California, Santa Barbara. He then worked as a developmental technician in High Energy Physics, which inspired him to continue his academic journey at Stanford University. At Stanford, he earned both an MS and PhD in Applied Physics, with a focus on astronomical detectors. His doctoral work explored the development of Hybrid CMOS SiPIN Detectors for use as astronomical imagers. Following his PhD, he contributed to the Atacama Large Millimeter/Submillimeter Array in Chile before transitioning to Lawrence Livermore National Laboratory (LLNL) in 2010. At LLNL, his research spans physics and engineering, with projects ranging from modeling gamma-ray propagation to developing firmware for sensors and FPGAs. He also consults with NASA on the James Webb Space Telescope, where he develops detector control firmware.
\end{biographywithpic}

\begin{biographywithpic}
{Peter Heatwole}{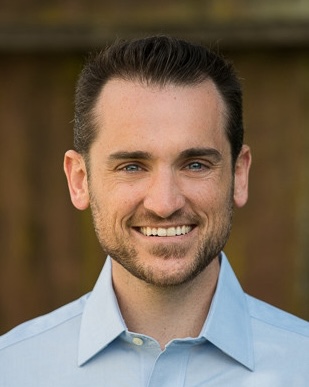} is an embedded software engineer at Lawrence Livermore National Laboratory. He graduated with a B.S. in Computer Engineering and an M.S in Electrical Engineering from California Polytechnic State University-San Luis Obispo. He works on the Pandora payload flight software for the Pandora Mission.
\end{biographywithpic}

\begin{biographywithpic}
    {Christina Hedges}{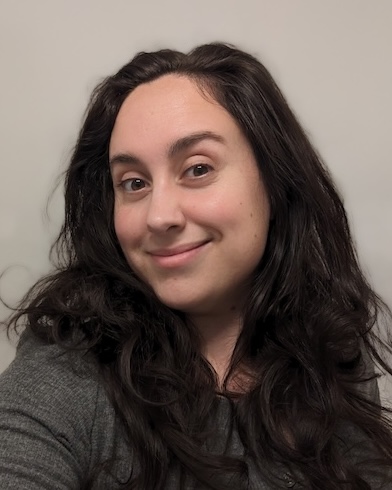}
    is a scientist based at NASA's Goddard Space Flight Center, where she leads the TESS Science Support Center (TSSC), and the lead pipeline scientist for NASA Pandora. She specializes in developing software tools and data analysis methods for NASA missions, with a focus on open science and supporting the broader astronomical community. She holds a PhD in astrophysics and has extensive experience in data analysis for space-based observatories. She is also passionate about promoting open-source software practices through her work in developing accessible tools for the scientific community.
\end{biographywithpic}

\begin{biographywithpic}
    {Jessie Dotson}{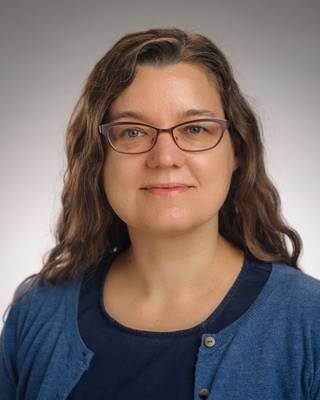}
    is the Deputy Principal Investigator of the Pandora SmallSat mission and an Astrophysicist in the Space Science and Astrobiology Division at NASA's Ames Research Center. She was previously the Project Scientist for NASA's Kepler Space Telescope. She earned a Ph.D. in astronomy and astrophysics from the University of Chicago. Her areas of specialization were infrared instrumentation, polarization measurements and the role of magnetic fields in giant molecular clouds. She has worked on the development of infrared and sub-millimeter polarimeters, spectrometers and cameras for ground-based, airborne and space-based observatories. She has served as the SOFIA Instrument Scientist, and she served six years as the Branch Chief for Astrophysics at NASA Ames Research Center. In 2016, she was awarded a NASA Outstanding Leadership Medal.
\end{biographywithpic}

\begin{biographywithpic}
    {Thomas Greene} {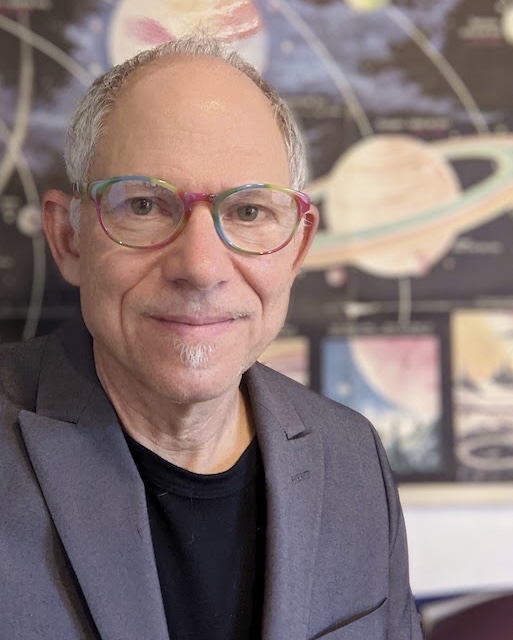} is an Astrophysicist in the Space Science and Astrobiology Division at NASA’s Ames Research Center. His research interests include observational studies of young stars and exoplanets as well as developing astronomical observatories and instrumentation. Greene was a co-investigator on the JWST NIRCam and MIRI JWST instruments and is now leading the MANATEE guaranteed time observing program to characterize exoplanets with those instruments. Before joining NASA, he was a staff scientist at the Lockheed Martin Advanced Technology Center, on the faculty of the University of Hawaii, and on the staff of the NASA Infrared Telescope Facility. Greene received a Ph.D. in astronomy from the University of Arizona.
\end{biographywithpic}

\begin{biographywithpic}
    {Trevor Foote}{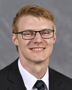}
    is currently a Ph.D. candidate in the Astronomy Department at Cornell University. He received his bachelor's degrees in civil engineering and astrophysics from Washington State University and his master’s degree in astronomy from Cornell University. His research interests include mission planning and detector development, specifically for exoplanet atmospheric characterization science.
\end{biographywithpic}

\begin{biographywithpic}
{Nikole Lewis}{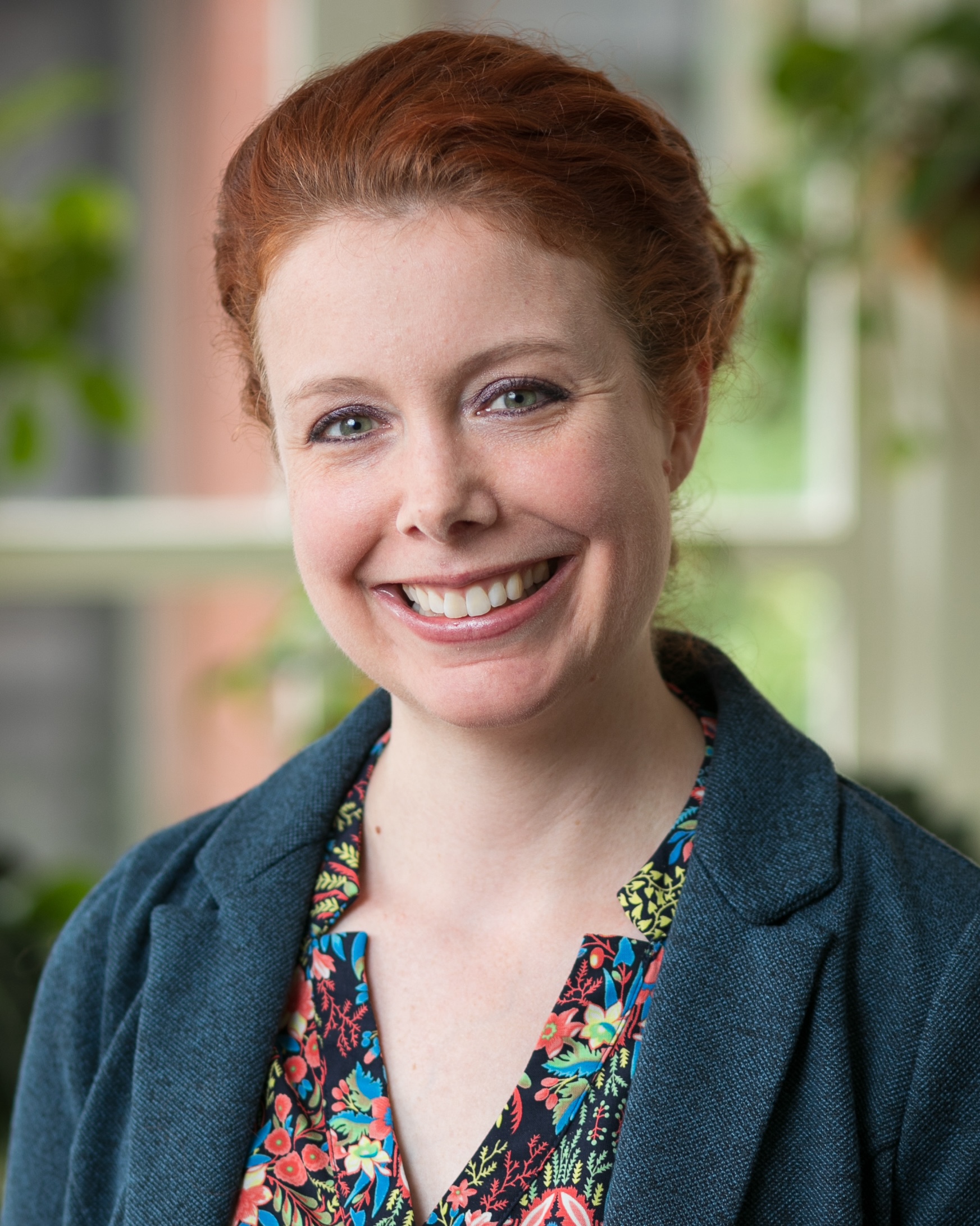} is an expert in the area of exoplanet characterization. She has successfully bridged the gap between theory and observation through her pioneering work with Spitzer, Hubble and JWST exoplanet observations and the development of associated modeling and analysis tools. Her current research includes work with both space- and ground-based campaigns to measure spectra and photometry of exoplanet atmospheres as well as the development of one-, two-, and three-dimensional atmospheric models to guide and interpret those observations.
\end{biographywithpic}

\begin{biographywithpic}
    {Benjamin Rackham}{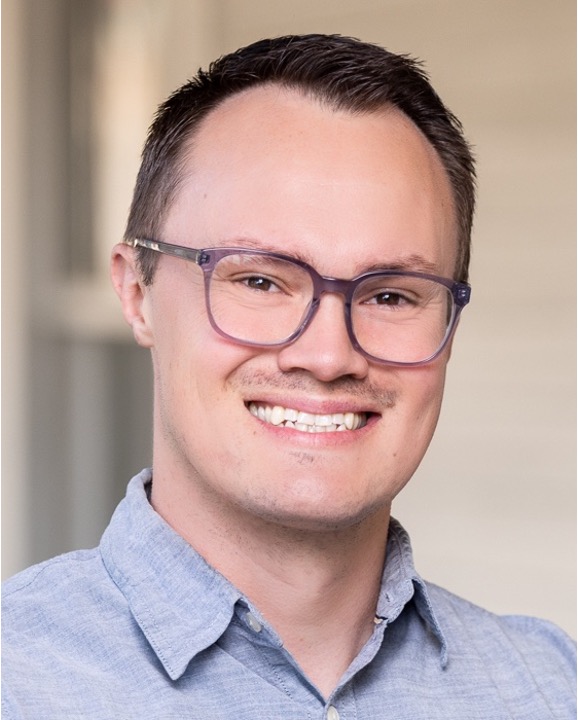}
    is a Research Scientist at MIT. He works on finding and characterizing transiting exoplanets, with an emphasis on understanding the activity of exoplanet host stars and how it impacts transit measurements. 
    Along with leading the Stellar Contamination Working Group for Pandora, he is the PI of HST and JWST Legacy Archival Programs with the goal of characterizing the stellar photospheric heterogeneity and activity of exoplanet host stars. 
    Before his current position, he received a PhD in Astronomy \& Astrophysics from the University of Arizona and was a 51 Pegasi b Postdoctoral Fellow at MIT.
\end{biographywithpic}

\begin{biographywithpic}
    {Brett Morris} {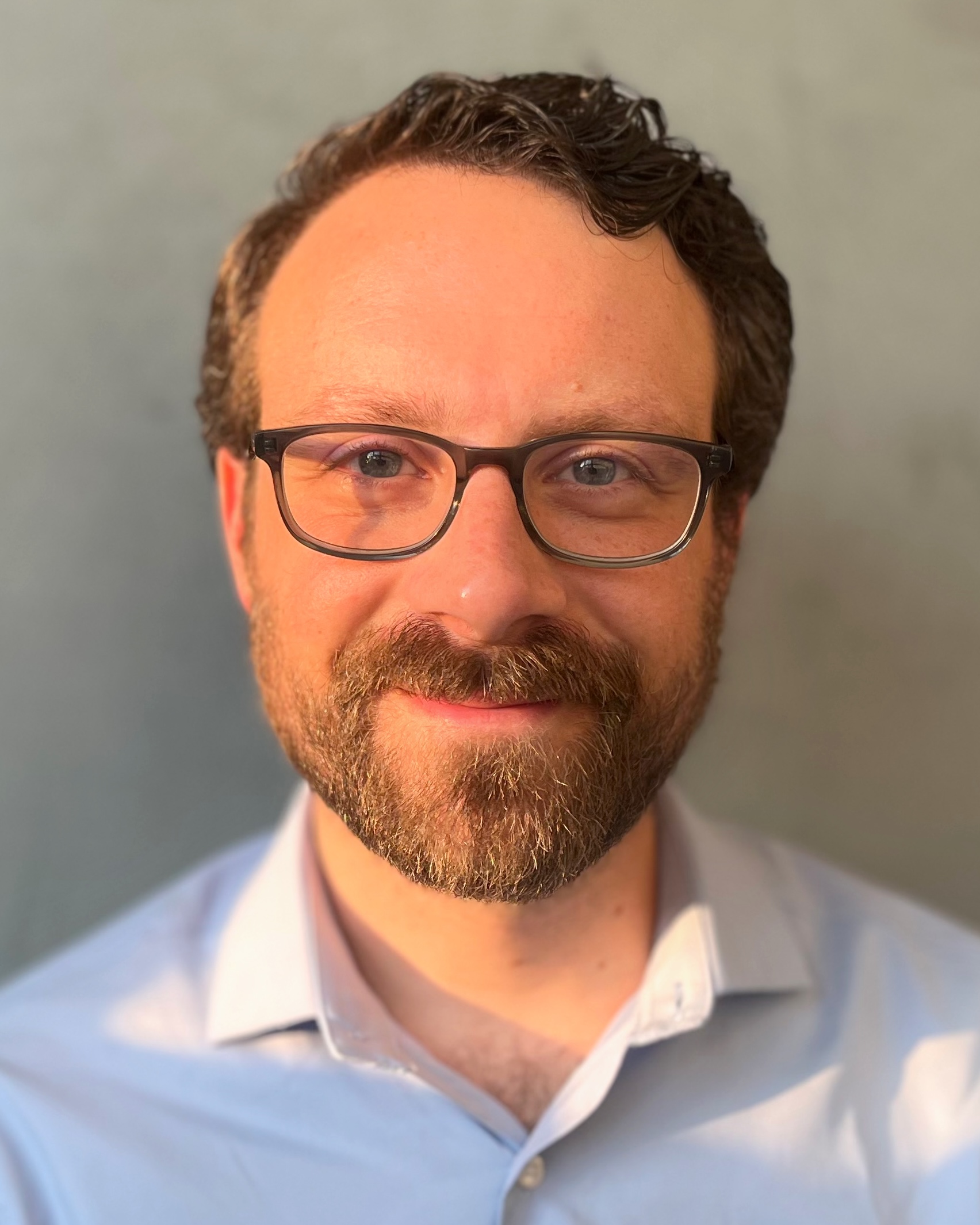} is a software engineer at the Space Telescope Science Institute in Baltimore, Maryland. He works on data analysis tools for JWST and the Roman Space Telescope, and leads research on characterization of exoplanet atmospheres while mitigating the effects of stellar magnetic activity. Morris received a Ph.D. in astronomy and astrobiology from the University of Washington.
\end{biographywithpic}

\begin{biographywithpic}
    {Emily Gilbert} {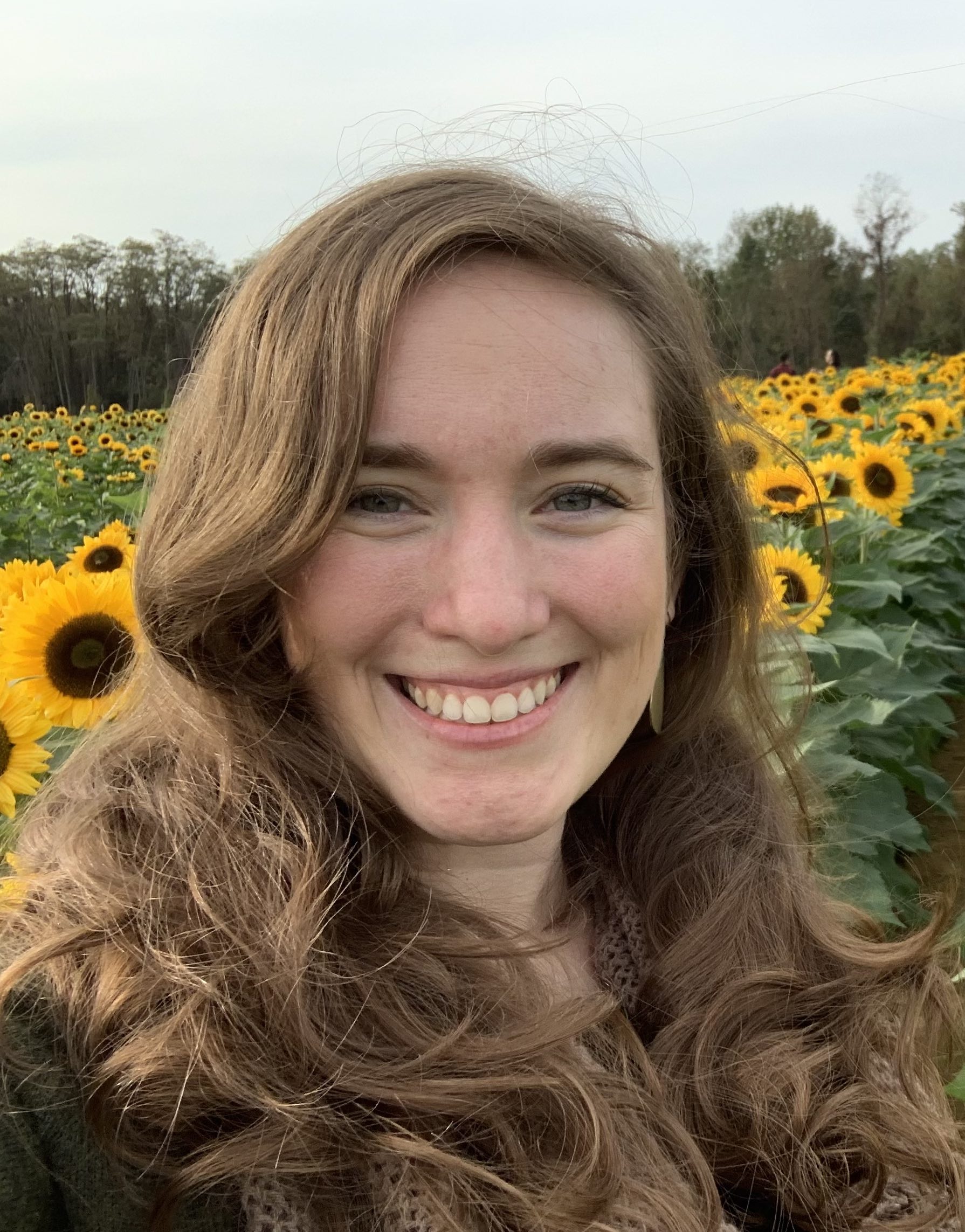} is a postdoctoral fellow at NASA's Jet Propulsion Laboratory. Emily's research focuses on the detection and characterization of exoplanets orbiting low mass stars using data from the Transiting Exoplanet Survey Satellite (TESS) mission. She received her Sc.B. in Astrophysics from Brown University and went to the University of Chicago for graduate school and completed her research with scientists at the Adler Planetarium and the TESS team at NASA Goddard Spaceflight Center. For her dissertation, Emily used TESS data to detect and characterize exoplanets orbiting M dwarf stars.
\end{biographywithpic}

\begin{biographywithpic}
    {Veselin Kostov}{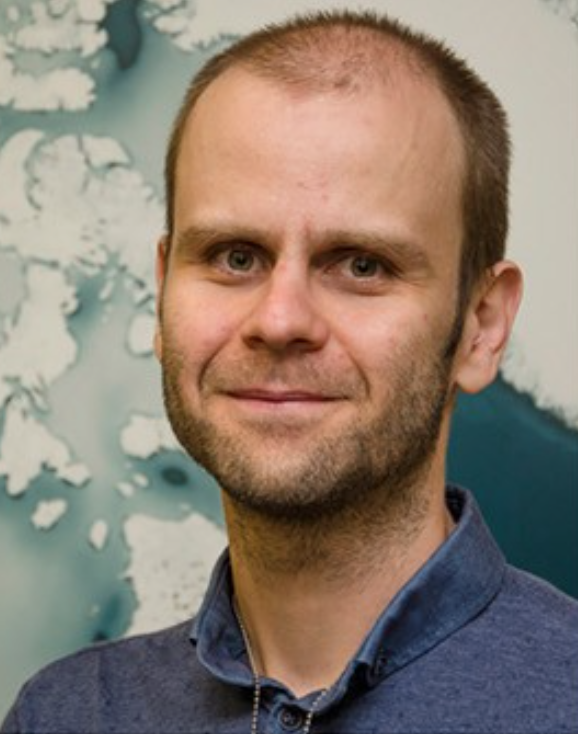} is a research scientist at NASA's Goddard Space Flight Center and SETI Institute. He is a member of Pandora's Science Operation Center and is responsible for the development, maintenance and application of Pandora's scheduling pipeline. Previously, he was a NASA Postdoctoral Fellow at Goddard Space Flight Center and also spent a year as a McLean Postdoctoral Fellow at University of Toronto's Department of Astronomy. He recieved his Ph.D.from the Johns Hopkins University Department of  Physics and Astronomy.
\end{biographywithpic}

\begin{biographywithpic}
{Jason Rowe}{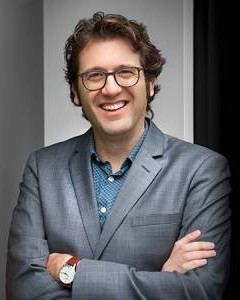} is the Canada Research Chair in Exoplanet Astrophysics and Professor in the Department of Physics and Astronomy at Bishop's University.  His current research goals are to determine what properties make a planet ‘Earth-like’ and whether there is life beyond Earth. He has authored and co-authored over 200 publications with over 20~000 total citations.  He is Principal Investigator of the Canadian POET mission to discover and characterize exoplanets.  He is a science team member of the MOST, Kepler, NEOSSat and CASTOR mission.  Rowe received a Ph.D. in astrophysics from the University of British Columbia.
\end{biographywithpic}

\begin{biographywithpic}
    {Lindsey Wiser}{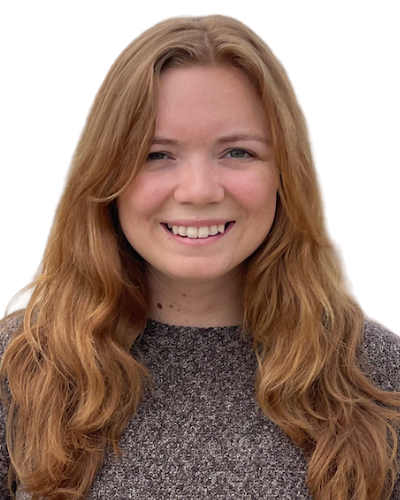} is an Astrophysics PhD Candidate at Arizona State University. Her research interests include exoplanet atmosphere modeling, transit spectroscopy with space telescopes, and astrophysics mission formulation. She earned her Bachelor of Science in 2020 from Johns Hopkins University studying engineering mechanics and Earth and planetary science. Lindsey is a student shadow with the Pandora science team working on planning instrument commissioning and community outreach initiatives. 
\end{biographywithpic}

\end{document}